\documentclass[aps,prl,showpacs,superscriptaddress,amsmath,amssymb,floatfix]{revtex4-1}
\usepackage{graphicx}
\usepackage{dcolumn}
\usepackage{bm}
\begin{document}

\title{Correlation function structure in square-gradient models of the liquid-gas interface: Exact results and reliable approximations.}

\author{A.O.\ Parry}
\affiliation{Department of Mathematics, Imperial College London, London SW7 2BZ, UK}

\author{C.\ Rasc\'{o}n}
\affiliation{GISC, Departamento de Matem\'aticas, Universidad Carlos III de Madrid, 28911 Legan\'es, Madrid, Spain}
\affiliation{ICMAT, Campus Cantoblanco UAM, 28049 Madrid, Spain}

\begin{abstract}
In a recent article, we described how the microscopic structure of density-density correlations in the fluid interfacial region, for systems with short-ranged forces, can be understood by considering the resonances of the local structure factor occurring at specific parallel wave-vectors $q$. Here, we investigate this further by comparing approximations for the local structure factor and correlation function against three new examples of analytically solvable models within square-gradient theory. Our analysis further demonstrates that these approximations describe the correlation function and structure factor across the whole spectrum of wave-vectors, encapsulating the cross-over from the Goldstone mode divergence (at small $q$) to bulk-like behaviour (at larger $q$). As shown, these approximations are exact for some square-gradient model potentials, and never more than a few percent inaccurate for the others. Additionally, we show that they very accurately describe the correlation function structure for a model describing an interface near a tricritical point. In this case, there are no analytical solutions for the correlation functions, but the approximations are near indistinguishable from the numerical solutions of the Ornstein-Zernike equation.
\end{abstract}



\maketitle

\section{Introduction}

The nature of density-density correlations at the liquid-gas interface has attracted enourmous attention since Buff, Lovett and Stilinger showed that, at long wavelengths, these correlations can be understood using a mesoscopic description where the area of the interface is resisted by the surface tension \cite{BLS1965,Zittartz1967,Wertheim1976,Weeks1977,Evans1979,Evans1981,Rowlinson1982,Aarts2004}. This capillary wave model of the interface, which regards it as a drumskin under tension, is an excellent description of the interfacial region for wavelengths much larger than the bulk liquid correlation length, and has been successfully used to understand fundamental interfacial phase transitions such as roughnening and wetting \cite{Dietrich1988,Schick1990,Forgacs1991}. However, the question of how density-density correlations behave at shorter lengthscales, comparable with the bulk correlation length, has proved considerably more difficult to answer \cite{Rochin1991,Napiorkowski1993,Parry1994,Robledo1997,Mecke1999,Fradin2000,Blokhuis2008,Blokhuis2009,Parry2014,Hofling2015,Chacon2016,Hernandez2016,Parry2016,Macdowell2017,Hernandez2018}. For instance, plausible attempts to extend the capillary wave description by introducing a scale-dependent surface tension have run into numerous difficulties and have failed to connect with detailed simulation studies of the correlation function $G$ and its integral, the structure factor $S$ \cite{Hofling2015}. 
In recent articles \cite{Parry2016,Parry2019}, we have shown that the properties of the correlation function and structure factor can be determined directly using the formalism of Density Functional Theory (DFT), without having to resort to extended mesoscopic theories. Indeed, we showed that this approach quantitatively explains the results of the largest simulation study of correlation functions near the liquid-gas interface in a system with truncated Lennard-Jones interactions, which as mentioned above are not consistent with mesoscopic approaches. In particular, we showed that the higher wave-vector behaviour of the structure factor is determined by a hierarchy of resonances occurring at specific values of the wavevector $q$, which further precise the connection between the structure factor $S$ and the underlying two-point correlation function $G$. The existence of these resonances constrains strongly the structure of these two functions, allowing us to put forward a family of robust aproximations for $G$ and $S$ across all wavevectors. The purpose of the present paper is to further check the validity of these approximations for a number of models, including some which are analytically solvable and have not been reported previously. 

Our paper is arranged as follows: In the first section, we recall the basics of the DFT formalism, focusing on square-gradient theory and the general relation between the structure factor and correlation function determined by the resonances. In the following section, we present exact analytical results for a number of models, which, in section 3, are compared with the aforementioned approximations. We also report results for correlation function structure for interfaces near a tricritical point. A summary and conclusions finish the paper.

\section{Formalism}

\subsection{DFT and Square-Gradient Theory}

Within Density Functional Theory, the equilibrium density profile, surface tension and correlation functions are obtained from the Grand Potential functional $\Omega[\rho]=F[\rho]-\int\!d{\bf r}\,(\mu-V({\bf r}))\,\rho({\bf r})$, where $\rho({\bf r})$ is the density distribution, $\mu$ is the chemical potential, and $V({\bf r})$ is the external potential \cite{Evans1979}. The equilibrium density profile is obtained from minimization of $\Omega[\rho]$
\begin{equation}\label{one}
      \frac{\delta\Omega[\rho]}{\delta\rho({\bf r})}=0
\end{equation}
while  the direct correlation function of the inhomogeneous fluid
\begin{equation}
      C({\bf r},{\bf r}')=\frac{1}{k_B T}\,\frac{\delta^2\,F[\rho]}{\delta\rho({\bf r})\delta\rho({\bf r}')}
\end{equation}
is obtained as the second derivative of the intrinsic Helmholtz Free-energy functional $F[\rho]$, which must be evaluated at the equilibrium fluid density. Hereafter, we use units in which $k_BT=1$. From $C({\bf r},{\bf r}')$, we can then determine the equilibrium density-density correlation function $G({\bf r}',{\bf r})$ via the solution of the Ornstein-Zernike equation
\begin{equation}
      \int d{\bf r}''\;C({\bf r},{\bf r}'')\,G({\bf r}'',{\bf r}')\;=\;\delta({\bf r}-{\bf r}')
\end{equation}
Consider the mean-field square-gradient theory based on the model Grand Potential functional \cite{Evans1979}
\begin{equation}
\Omega[\rho]=\int\!\! d{\bf{r}}\,\left(\frac{f}{2}(\nabla \rho)^2+\Delta\phi(\rho)\right)
\label{SGT}
\end{equation}
where, for simplicity, we set $f=1$, since this does not appear in our final results. This is the simplest microscopic theory of the interfacial region, applicable to systems with short-ranged forces. Although it does not allow for dispersion forces, the capillary wave broadening of the interface or packing effects at a molecular scale, it still provides invaluable insight into interfacial properties.

Below the critical temperature $T_c$, the bulk free-energy density $\phi(\rho)$ has a standard double well structure describing the coexistence of liquid and gas phases with densities $\rho_l$ and $\rho_g$, respectively, for which $\phi(\rho_l)=\phi(\rho_g)$. The shifted potential $\Delta\phi(\rho)\equiv\phi(\rho)-\phi(\rho_b)$ simply subtracts a bulk contribution. The second derivatives of the potential $\phi''(\rho_b)=\kappa_b^2$ then determine the inverse correlation length $\kappa_b\equiv 1/\xi_b$ of the bulk liquid ($b=l$) or gas ($b=g$) phase. These characterise the exponential decay of the bulk correlation function $G_b(r)$ (where $r$ is the intermolecular separation) the three-dimensional Fourier transform of which identifies the bulk factor
\begin{equation}
 S_b(q)\;=\;\frac{S_b(0)}{\,1+\xi_b^2q^2\,}
 \label{BulkS}
\end{equation}
where $S_b(0)=(\Delta \phi''(\rho_b))^{-1}$ is the bulk compressibility. It is also convenient to introduce the 2D Fourier transform of $G_b(r)$ along the $x$-$y$ plane, which is determined as 
\begin{equation}
G_b(z;q)\;=\;\frac{1}{2\kappa_q}\,e^{-\kappa_q z}
\end{equation}
where $\kappa_q\equiv\sqrt{\kappa_b^2+q^2}$. As we are going to concentrate on systems with an Ising symmetry, we drop the subscript $b$ hereafter, except for $S_b(q)$ and $G_b(z;q)$, in order to emphasise that these are bulk functions.

We suppose that a planar interface of macroscopic area separates the bulk phases near 
the $z=0$ plane. The equilibrium density profile $\rho(z)$ is determined by (\ref{one}) and satisfies the Euler-Lagrange equation
\begin{equation}
\frac{d^2\rho}{dz^2}=\Delta\phi'(\rho)
\label{EL}
\end{equation}
subject to boundary conditions $\rho(\infty)=\rho_l$ and $\rho(-\infty)=\rho_g$. This has the first integral determining that
\begin{equation}
\rho'(z)=\sqrt{2\Delta\phi}
\label{first}
\end{equation}
and leads to the famous van der Waals formula for the surface tension
\begin{equation}
\sigma=\int_{-\infty}^\infty\!\!\!\! dz\,\; \rho '(z)^2
\label{st}
\end{equation}
where more generally there is a pre-factor $f$, which, recall, we have set to $1$.
For this model, the direct correlation function reduces to
\begin{equation}
      C({\bf r},{\bf r}')=\left(-\nabla^2_{{\bf r}}+\phi''(\rho({\bf r}))\right)\,\delta({\bf r}-{\bf r}')
\end{equation}
and, therefore, its 2D Fourier transform along the interface is the delta function operator
\begin{equation}
C(z,z';q)=\left(-\partial^2_z+q^2+\phi''(\rho(z))\right)\,\delta(z-z')
\end{equation}

Thus, the Ornstein-Zernike (OZ) equation for the 2D Fourier transform of the density-density correlation function $G$ reduces to the differential equation
\begin{equation}
\Big(-\partial^2_z+q^2+\phi''\big(\rho(z)\big)\Big)\;G(z,z';q)\;=\;\delta(z-z')
\label{OZGTG}
\end{equation}
The local structure factor is defined as the integral
\begin{equation}
      S(z;q)=\int_{-\infty}^\infty\!\!\!\!dz'\;\,G(z,z';q)
\end{equation}
and, therefore, satisfies
\begin{equation}
\left(-\partial^2_z+q^2+\phi''(\rho(z))\right)\,S(z;q)=1
\label{OZGTS}
\end{equation}

\subsection{Five Properties of $S$ and $G$}

{\bf A) Wertheim-Weeks Goldstone mode.} In the limit of $q\to 0$, the structure factor necessarily has a Goldstone mode divergence
\begin{equation}
S(z;q)\;=\; \frac{\;\Delta\rho\,\rho'(z)\;}{\sigma q^2}\;+\;\cdots
\label{GM}
\end{equation}
where $\Delta\rho\equiv\rho_l\!-\!\rho_g$, and the higher-order terms are of order $q^0$. This result, which is consistent with an exact sum-rule analysis due to Wertheim-Weeks \cite{Wertheim1976} and the expectations of capillary-wave theory \cite{Weeks1977,Parry2016}, follows directly from the spectral expansion of the two-point function
\begin{equation}
G(z,z'q)\;=\;\sum_n \frac{\;\psi^*_n(z)\,\psi_n(z')\,}{E_n+q^2}
\label{spectral}
\end{equation}
as first shown by Evans \cite{Evans1979}. Here, the eigenfunctions satisfy the Schr\"odinger-like equation 
\begin{equation}
\Big(\!-\partial_z^2+\Delta\phi''\big(\rho(z)\big)\Big)\,\psi_n(z)\;=\;E_n\,\psi_n(n)
\label{OZschrod}
\end{equation}
The low $q$ divergence (\ref{GM}) then follows from noting that the normalised ground-state $\psi_0(z)\propto\rho'(z)$ has zero energy ($E_0=0$) by virtue of the Euler-Lagrange equation (\ref{EL}). That is the correlation function must contain the small $q$ divergence 
\begin{equation}
G(z,z';q)\;=\; \frac{\;\rho'(z)\,\rho'(z')\;}{\sigma q^2}\;+\;\cdots
\label{GMG}
\end{equation}
integration of which gives (\ref{GM}).\\

{\bf B) Integral Sum-Rule.} Multiplying equation (\ref{OZGTS}) by $\rho'(z)$ and integrating over all $z$, we find
\begin{equation}
\int_{-\infty}^\infty\!\!\!\! dz\;\, \rho'(z)\Big(-\partial_z^2+q^2+\Delta\phi''\big(\rho(z)\big)\Big)\, S(z;q)\;=\;\Delta\rho
\end{equation}
which, on using the Euler-Lagrange equation (\ref{EL}), reduces to
\begin{equation}
\int_{-\infty}^\infty\!\!\!\! dz\;\, \rho'(z)\, S(z;q)\;=\;\frac{\,\Delta\rho\,}{q^2}
\label{Sumrule}
\end{equation}
which, more generally, contains a factor $f$ in the denominator. Thus, any complicated wave-vector dependence present in $S(z;q)$ may be eliminated by taking a weighted integral over the interfacial region, leaving only a pure Goldstone mode. This result can also be obtained from the spectral expansion (\ref{spectral}), noting that $\rho'(z)$ is proportional to the lowest-order eigenfunction $\psi_0(z)$.\\

{\bf C) Large $z$ decay.} At fixed $q$, the structure factor either side of the interface decays towards its bulk liquid or gas value as $z\to |\infty|$ according to
\begin{equation}
S(z;q)\;=\; S_b(q)\;+\;\frac{\Delta\rho\,\rho'(z)}{\,\sigma_1\, q^2(1+\xi^2q^2)\,}\;+\;\cdots
\label{PRE}
\end{equation}
where the higher-order terms are of order $\mathcal{O}\big(e^{-\kappa_q|z|},e^{-2\kappa|z|}\big)$ and decay faster than $\rho'(z)\propto e^{-\kappa |z|}$ \cite{Parry2016}. The surface tension-like coefficient appearing in the denominator of (\ref{PRE}) (referred to as $\sigma_b(0)$ in \cite{Parry2016}) is given by 
\begin{equation}
\sigma_1\;=\;\kappa\,\Delta\rho\,\frac{\Delta \phi''(\rho_b)}{\,|\Delta\phi'''(\rho_b)|\,}
\label{sigma1}
\end{equation}
and is determined entirely by the appropriate bulk liquid or gas quantities. Thus, in the absence of a perfect Ising symmetry, the coefficient $\sigma_1$ is {\it different} either side of the interface. In addition, even if Ising symmetry is present, the coefficient $\sigma_1$ is not equal to the surface tension in general, although intriguingly this is true for the standard quartic potential (see below). The crucial insight from the result (\ref{PRE}) is that the whole wave-vector dependence of the leading-order exponentially decaying term is related directly to the bulk structure factor. Indeed, it arises explicitly from the combination $\,S_b(q)S_b(i\sqrt{\kappa_b^2-q^2})\propto 1/q^2( 1+\xi^2q^2)\,$ \cite{Parry2016}.\\

{\bf D) Resonances.} The asymptotic expression (\ref{PRE}) appears to contain a Goldstone mode divergence as $q\to 0$ but is not consistent with the Wertheim-Weeks result (\ref{GM}) since in general $\sigma_1\ne\sigma$. This discrepancy is explained by noting that, in the limit $q\to 0$, the terms of order $\mathcal{O}\big(e^{-\kappa_q|z|}\big)$ appearing in $S(z;q)$ must also be included, in order to capture the full divergence correctly. These themselves can be related to a hierarchy of resonances that occur at specific wavevectors $q=\sqrt{3}\kappa$, $\sqrt{8}\kappa$, $\sqrt{15}\kappa$, $\cdots$ Provided the potential $\phi(\rho)$ has an analytic expansion about the bulk density, these determine that the local structure factor throughout the interfacial region can be written as \cite{Parry2019}
\begin{equation}\label{Sres}
S(z;q)=S_b(q)+\frac{\Delta\rho\,\rho'(z)}{\sigma_1\,q^2(1+\xi^2q^2)}+\frac{\Delta\rho}{\rho'(0)}\sum_{n=2}^\infty\;\frac{\sigma}{\sigma_n}\;\frac{ G(0,z;q)-G(0,z;\sqrt{n^2-1}\kappa)}{(1+\xi^2 q^2)(1-\frac{\xi^2 q^2}{n^2-1})}
\end{equation}
In this expression, the origin $z=0$ is chosen to be at the point where $\rho'(z)$ is maximum.
The resonances are weighted by generalised surface tension-like coefficients $\sigma_n$, which can be determined from the correlation function, satisfying the relation
\begin{equation}
\frac{1}{\sigma}=\frac{1}{\sigma_{1}}+\frac{1}{\sigma_{2}}+\frac{1}{\sigma_{3}}+\cdots
\label{sigsum}
\end{equation}
This ensures that, in the limit $q\to 0$, the local structure factor exhibits the required Goldstone mode divergence. It is also straightforward to show that the expression (\ref{Sres}) satisfies the integral sum-rule (\ref{Sumrule}).\\

{\bf E) Reliable Approximations.} The additional relation between $S$ and $G$ provided by the resonances, Eq.~(\ref{Sres}), constrains strongly the properties of both functions and provides a scheme for classifying different model potentials according to the presence (or absence) of specific resonances. It also leads to new classes of fully integrable models for which the density profile, surface tension, and correlation functions can be determined analytically. These will be discussed in the following section.
Perhaps more importantly, it also points towards a very robust approximation applicable to all potentials for the structure factor at the origin. To this end, let us suppose that $G(0,0;q)\approx\rho'(0)^2/\sigma q^2 + \mathcal{C}$, containing the Goldstone mode and an unknown correction term which we approximate as a constant $\mathcal{C}$. Subtitution into (\ref{Sres}) leads to
\begin{equation}
\frac{S(0;q)}{S_b(q)}\;\approx\; 1 + \frac{\Delta\rho\rho'(0)}{\sigma q^2\,S_b(0)}
\label{Sapp}
\end{equation}
which is independent of the constant $\mathcal{C}$ \cite{Parry2019}. Indeed, this expression is the exact result for the quartic Landau potential \cite{Parry2016}.

Similarly, an analysis of the correlation function structure for different model potentials reveals that the correlation function at the origin is always well approximated by 
\begin{equation}
\frac{G(0,0;q)}{G_b(0;q)}\;\approx\;\, 1\;+\;\frac{\rho'(0)^2}{\sigma q^2G_b(0;0)}
\label{Gapp}
\end{equation}
which is also the exact result for the quartic Landau potential \cite{Parry2014}. Thus, at the origin (and only at the origin), the approximate rule of factoring out a bulk background {\it and} a multiplicative correction to the Goldstone mode describes accurately the correlation function and structure factor over the whole range of wave-vectors. In the next section, we test these approximations against examples of fully integrable model potentials, only some of which have been reported recently.\\

\section{Fully Integrable Models}

\subsection{Landau Quartic Potential}

\begin{figure}[t]
\includegraphics[width=0.5\columnwidth]{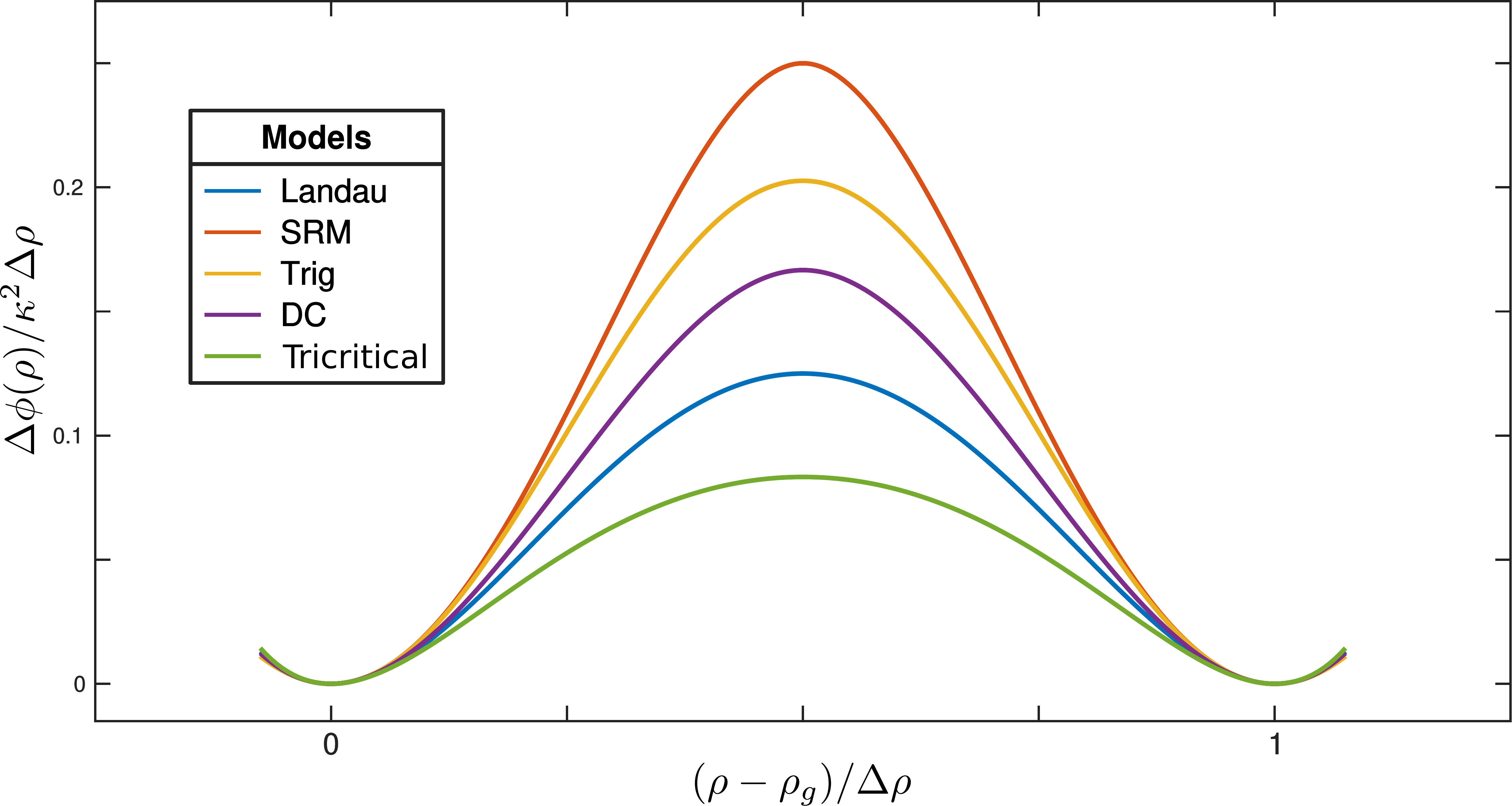}
\caption{\label{Fig1} Potentials $\Delta\phi(\rho)$ for the five models considered here: Landau quartic potential (Eq.~(\ref{m4})), Single-resonance model for $n=2$ (SRM) (Eq.~(\ref{SRMPhi2})), Trigonometric model (Eq.~(\ref{TM})), Double-cubic (DC) model (Eq.~(\ref{DC})), and Tricritical or Fisk-Widom model (Eq.~(\ref{FW})).
}
\end{figure}

Before we discuss new examples of analytically solvable square-gradient theories, we recall the results for the Landau quartic potential \cite{Zittartz1967,Parry2014}
\begin{equation}
\phi(\rho)=-\frac{t}{2}(\rho-\rho_c)^2+\frac{u}{4}(\rho-\rho_c)^4
\label{m4}
\end{equation}
Here $\rho_c=(\rho_l+\rho_g)/2$ is the critical density, $t\propto T_c-T$ and $u>0$ is a positive constant. We can also write this as:
\begin{equation}
\Delta\phi(\rho)\;=\;\frac{\;\kappa^2}{2}\,\frac{\,(\rho-\rho_l)^2\,(\rho-\rho_g)^2}{\Delta\rho^2}
\label{m42}
\end{equation}
where the dependence on the correlation length is explicit (see Fig.\ref{Fig1}). The solution of the Euler-Lagrange equation (\ref{EL}) leads to the celebrated result for the profile \cite{Evans1979,Rowlinson1982}
\begin{equation}
\rho(z)\;=\;\rho_c\,+\,\frac{\Delta\rho}{2}\tanh\left(\frac{\kappa z}{2}\right)
\label{m4rho}
\end{equation}
and, via (\ref{st}), to the surface tension
\begin{equation}
\sigma\;=\;\frac{\;\kappa\Delta\rho^2}{6}
\label{m4sigma}
\end{equation}
The OZ equation (\ref{OZGTS}) reads
\begin{equation}
\left(-\partial^2_z+q^2+\kappa^2-\frac{\,3\kappa^2}{2}\,\text{sech}^2\left(\frac{\,\kappa z}{2}\right)\right)\,S(z;q)\;=\;1
\end{equation}
which has the solution \cite{Parry2014}
\begin{equation}
S(z;q)\;=\; S_b(q)\,+\,\frac{\Delta\rho\,\rho'(z)}{\,\sigma q^2(1+\xi^2q^2)}
\label{m4S}
\end{equation}
A remarkable feature of the quartic potential is that the coefficient $\sigma_1$ describing the asymptotic decay of $S(z;q)$, Eq.~(\ref{sigma1}), is identical to the surface tension
\begin{equation}
\sigma_1\;=\;\frac{\;\kappa\Delta\rho^2}{6}
\end{equation}
with all other coefficients $\;\sigma_2=\sigma_3=\cdots=\sigma_n=\infty$, implying that the resonances are absent. It also follows that the approximation (\ref{Sapp}) for the local structure factor $S(0;q)$ is exact. 

For this model, the OZ equation for the correlation function (\ref{OZGTG}) is
\begin{equation}\label{m4OZ}
\left(-\partial^2_z+q^2+\kappa^2-\frac{\,3\kappa^2}{2}\,\text{sech}^2\left(\frac{\,\kappa z}{2}\right)\right)\,G(z,z';q)\;=\;\delta(z-z')
\end{equation}
which, ordering the positions $z>z'$, has the solution \cite{Zittartz1967}
\begin{equation}\label{m4Gzz}
G(z,z';q)\;=\;\alpha(q)\;\,\psi_-(z;q)\,\psi_+(z';q)
\end{equation}
where
\begin{equation}
      \begin{array}{ccl@{\,}l}
\psi_-(z;q) & = & e^{-\kappa_q z} & \Big(\tanh^2\left(\frac{\kappa z}{2}\right)+2\,\xi\kappa_q\tanh\left(\frac{\kappa z}{2}\right)+1+\frac{4}{3}q^2\xi^2\Big)\\[.35cm]
\psi_+(z;q) & = & e^{\kappa_q z} & \Big(\tanh^2\left(\frac{\kappa z}{2}\right)-2\,\xi\kappa_q\tanh\left(\frac{\kappa z}{2}\right)+1+\frac{4}{3}q^2\xi^2\Big)
      \end{array}
\end{equation}
and
\begin{equation}
\alpha(q)\;=\;\displaystyle\frac{3\,\kappa^2}{8\,\kappa_q\,q^2\left(1+\frac{4}{3}q^2\xi^2\right)}
\end{equation}
When both particles are at the origin, the expression for the correlation function simplifies to 
\begin{equation}
G(0,0;q)=\frac{3+4\,q^2\xi^2}{8\,q^2\xi^2\kappa_q}
\end{equation}
which can be rewritten as
\begin{equation}
G(0,0;q)=\frac{1}{2\kappa_q}+\frac{\rho'(0)^2}{\sigma q^2\sqrt{1+q^2\xi^2}}
\label{m4G}
\end{equation}
showing that the approximation (\ref{Gapp}) is exact.

\subsection{A Model with a Single Resonance}

The quartic potential (\ref{m4}) is a special example of a model that generates no resonances in $S(z;q)$. The natural generalization of this is the class of models which generate just a single resonance occurring at, say, $q=\sqrt{n^2-1}\,\kappa$. That is, models for which $\sigma_n=\sigma$ at one specific value of $n$, with all other weights $\sigma_{m}=\infty$ for $m\ne n$. In this case, the structure factor would have the form
 \begin{equation}
S(z;q)\;=\;S_b(q)+\frac{\Delta\rho}{\rho'(0)}\frac{\;G(0,z;q)-G(0,z;\sqrt{n^2-1}\,\kappa)}{(1+q^2\xi^2)(1-\frac{q^2\xi^2}{n^2-1})}
\label{SRMS}
\end{equation}
for that particular value of $n$ we have chosen \cite{Parry2019}. Substituting this expression into the Ornstein-Zernike equation (\ref{OZGTS}) for $S(z;q)$ determines that it is indeed a solution, provided that the two-point function satisfies \cite{Parry2019}
\begin{equation}
G(0,z;\sqrt{n^2-1}\,\kappa)\;\propto\; \phi''(\rho(z))-\kappa^2
\label{rq}
\end{equation}
It follows that the potentials $\phi(\rho)$ that display single resonances satisfy the non-linear fourth order equation
\begin{equation}
2\Delta\phi\, \phi ''''+\phi'\phi'''\;=\; (\phi''-\kappa^2)(\phi''+\kappa^2(n^2-1))
\label{SRMPhi}
\end{equation}
with boundary conditions $\,\Delta\phi(\rho_l)=\phi'(\rho_l)=\phi'(\rho_c)=0\,$ and $\,\phi''(\rho_l)=\kappa^2\,$.\\
For $n=1$, the solution of (\ref{SRMPhi}) recovers the Landau quartic potential, and eq.~(\ref{SRMS}) reduces to (\ref{m4S}), as can be seen directly by analytic continuation and taking the limit $n\to 1$.\\

For $n=2$, solution of (\ref{SRMPhi}) leads to 
\begin{equation}
\Delta\phi(\rho)\;=\;
\left\{
\begin{array}{ll}
\displaystyle\frac{\;\kappa^2}{2}(\rho-\rho_g)^2 \left(1-2\,\frac{(\rho-\rho_g)^{2}}{\Delta\rho^{2}}\right) & \text{for}\;\;\rho_g\le\rho\le\rho_c \\[.5cm]
\displaystyle\frac{\;\kappa^2}{2}(\rho-\rho_l)^2\left((1-2\,\frac{(\rho-\rho_l)^{2}}{\Delta\rho^{2}}\right) & \text{for}\;\;\rho_c\le\rho\le\rho_l
\end{array}\right.
\label{SRMPhi2}
\end{equation}
which is continuous and differentiable at $\rho=\rho_c$ (see Fig.\ref{Fig1}). For this model potential, the local structure factor has the form 
 \begin{equation}
S(z;q)\;=\;S_b(q)+\frac{\Delta\rho}{\rho'(0)}\,\frac{\,G(0,z;q)-G(0,z;\sqrt{3}\kappa)}{(1+q^2\xi^2)(1-\frac{q^2\xi^2}{3})}
\label{SRMS2}
\end{equation}
showing a single resonance at $q=\sqrt{3}\,\kappa$ together with a Goldstone mode divergence as $q\to 0$ arising implicitly from the singularity in the two-point function. This then further reduces to 
 \begin{equation}\label{SRMSzq}
S(z;q)\;=\;S_b(q)+\sqrt{8}\,\xi\;\frac{\,G(0,z;q)-G(0,z;\sqrt{3}\kappa)}{(1+q^2\xi^2)(1-\frac{q^2\xi^2}{3})}
\end{equation}
on using the result $\rho'(0)=\kappa\Delta\rho/\sqrt{8}$, which follows from (\ref{first}).\\
The density profile and surface tension for the potential (\ref{SRMPhi2}) can be determined analytically. For example, direct integration of (\ref{first}) yields, 
\begin{equation}
\rho(z)\;=\;\rho_b\,\mp\,\Delta\rho\,\frac{(2+\sqrt{2})e^{-\kappa |z|}}{\,3+2\sqrt{2}+e^{-2\kappa |z|}}
\label{rhotoy}
\end{equation}
with the minus (plus) sign applying on the liquid (gas) side of the interface, while for the surface tension we find
\begin{equation}
\sigma\;=\;\frac{1}{3}\left(1-\frac{1}{2\sqrt{2}}\right)\kappa\,\Delta \rho^2
\label{sigmatoy}
\end{equation}
which follows from (\ref{st}).\\

We can continue further and determine results for both the  pair-correlation function and structure factor at the resonant wave-vector. From (\ref{rq}), it follows that the correlation function must be given by 
\begin{equation}
G(0,z;\sqrt{3}\,\kappa)\;=\;\frac{\kappa^2-\phi''(\rho(z))}{\,2\,\rho'(0)\,\Delta\phi'''(\rho_c)}
\end{equation}
where the value of the amplitude follows from simply applying the boundary condition $\partial_z G(0,z=0^+;q)=-1/2$. Substituting the potential (\ref{SRMPhi2}) and density profile (\ref{rhotoy}), this yields explicitly
\begin{equation}
G(0,z;\sqrt{3}\,\kappa)=\frac{\sqrt{2}}{\kappa}\left(\frac{2+\sqrt{2}}{3+2\sqrt{2}}\right)^2\frac{e^{-2\kappa |z|}}{\left(1+\frac{1}{3+2\sqrt{2}}\,e^{-2\kappa|z|}\right)^2}
\label{Grestoy}
\end{equation}

Integration of the correlation function then determines that, exactly at the resonant wave-vector, the local structure at the interface takes the value 
\begin{equation}
S(0;\sqrt{3}\,\kappa)\;=\; \left(\frac{2+\sqrt{2}}{4+3\sqrt{2}}\right) \xi^2
\label{Srestoy}
\end{equation}
Thus, even at this relatively large wave-vector, the structure factor at the interface is still nearly double its bulk value $S_b(\sqrt{3}\kappa)=\xi^2/4$.\\

The above results for $G(0,z;\sqrt{3}\,\kappa)$ and $S(0;\sqrt{3}\kappa)$ are strongly suggestive that the single-resonance model for $n=2$ is fully integrable. Indeed, this is the case. Consider, for example, the OZ equation for $G(z,z';q)$, which reads
\begin{equation}\label{SRM2OZ}
\Big(-\partial^2_z+q^2+\kappa^2-6\kappa^2\,\text{sech}^2\big(\kappa (|z|-z_0)\big)\Big)\,G(z,z';q)\;=\;\delta(z-z')
\end{equation}
where $e^{2\kappa z_0}=1/(3+2\sqrt{2})$ or, equivalently, $\tanh(\kappa z_0)=-\sqrt{2}/2$. The similarity with the OZ equation for the Landau quartic potential (\ref{m4OZ}) suggests that we could try to find a similar solution. In particular, we seek solutions of (\ref{SRM2OZ}) which are an exponential $e^{\pm\kappa_q z}$ multiplied by a polynomial in $\tanh\big(\kappa(|z|-z_0)\big)$ (which turns out to be of second order). This leads to the full analytical solution for the correlation function. If $z\ge z'\ge 0$ (i.e. the particles are on the liquid side), we find that
\begin{equation}\label{SRMG1}
G(z,z';q)\;=\;\alpha(q)\;\psi_-(z;q)\,\psi_+(z';q)\;+\;\beta(q)\;\psi_-(z;q)\,\psi_-(z';q)
\end{equation}
while, if $z\ge 0\ge z'$ (i.e. they are on opposites sides of the interface),
\begin{equation}\label{SRMG2}
G(z,z';q)\;=\;\gamma(q)\;\,\overline{\psi}_+(z';q)\,\psi_-(z;q)
\end{equation}
where
\begin{equation}
      \begin{array}{ccl@{\,}l}
\psi_-(z;q) & = & e^{-\kappa_q z} & \Big(\tanh^2\big(\kappa(z-z_0)\big)+\xi\,\kappa_q\tanh\big(\kappa(z-z_0)\big)+\frac{1}{3} q^2\,\xi^2\Big)\\[.35cm]
\psi_+(z;q) & = &  e^{\kappa_q z} & \Big(\tanh^2\big(\kappa(z-z_0)\big)-\xi\,\kappa_q\tanh\big(\kappa(z-z_0)\big)+\frac{1}{3} q^2\,\xi^2\Big)\\[.35cm]
 \overline{\psi}_+(z;q) & = & e^{\kappa_q z} & \Big(\tanh^2\big(\kappa(z+z_0)\big)-\xi\,\kappa_q\tanh\big(\kappa(z+z_0)\big)+\frac{1}{3} q^2\,\xi^2\Big)
      \end{array}
\end{equation}
 The amplitudes of the correlation function are determined by
 \begin{equation}
\begin{array}{ccl}
\alpha(q)&=&\displaystyle\frac{9}{2\,q^2\xi^2(q^2\xi^2-3)\kappa_q}\\[.5cm]
\beta(q)&=&\displaystyle\frac{-3\,\alpha(q)}{(3+\sqrt{2}\,\xi\kappa_q)(3+2\,\xi^2q^2+3\sqrt{2}\,\xi\kappa_q)}\\[.5cm]
 \gamma(q)&=&\displaystyle\frac{9\,\sqrt{2}}{\xi\,q^2(3+\sqrt{2}\,\xi\kappa_q)(3+2\,\xi^2q^2+3\sqrt{2}\,\xi\kappa_q)}
\end{array}
\end{equation}

When both particles are at the origin, the expression for the correlation function simplifies to 
\begin{equation}
G(0,0;q)\;=\;\frac{3+2\,q^2\xi^2+3\sqrt{2}\,\xi\kappa_q}{2\,\xi\,q^2\left(3\,\sqrt{2}+2\,\xi\kappa_q\right)}
\label{SRMG00}
\end{equation}
which will be compared with the approximation (\ref{Gapp}) below.\\

Integration of $G(z,z';q)$, eqs.~(\ref{SRMG1}) and (\ref{SRMG2}), over $z'$ determines the local structure factor $S(z;q)$. However, this can be obtained much more easily by substituting $G(0,z;q)$ into expression (\ref{SRMSzq}), which clearly shows the resonance at $q=\sqrt{3}\,\kappa$.
Substituting (\ref{SRMG00}) into (\ref{SRMSzq}), at $z=0$, determines the exact value for the structure at the origin
\begin{equation}\label{SRMS0}
S(0;q)\;=\;\frac{6+2\,\xi^2\,q^2+3\sqrt{2}\,\xi\,\kappa_q}{\;q^2\,\left(2+2\,\xi^2\,q^2+3\sqrt{2}\,\xi\,\kappa_q\right)}
\end{equation}
which, at fixed $q$, is the maximum value of $S(z;q)$ in the interfacial region.\\

Finally, we mention that the potentials $\phi(\rho)$ obtained from solving (\ref{SRMPhi}) can be determined for other values of $n$, although their Taylor's expansions about each bulk density no longer truncates. However, it can be shown that these have a scaling form $\Delta\phi(\rho)=\frac{\kappa^2}{2}(\rho\!-\!\rho_l)^2\,W(x)$, where $x\equiv\alpha\big((\rho\!-\!\rho_l)/\Delta\rho\big)^{n}$, and the scaling function $W(x)$ has the expansion 
\begin{equation}
W(x)\;=\;1\,+\,x\,
-\,\frac{(n^2-4)}{4}\;x^2\,-\,\frac{(n^2-1)(n^2-4)}{48\,n^4}\;x^3\;+\;\cdots
\label{SRMn}
\end{equation}
which is valid for $n\ge 1$. The value of the pure number $\alpha$ can then be determined from the condition that $\phi'(\rho_c)=0$. The potential $\Delta\phi(\rho)$ truncates at quartic order for $n=1$ and $n=2$, consistent with the explicit results discussed above. 
In the limit of $n\to \infty$, the single-resonance potential approaches the shape of a simple double-parabola
\begin{equation}\label{DP}
      \Delta\phi(\rho)=
\left\{
\begin{array}{ll}
 \displaystyle\frac{\;\kappa^2}{2}\,(\rho-\rho_l)^2 & \textup{for} \;\;\rho>\rho_c\\[.5cm]
 \displaystyle\frac{\;\kappa^2}{2}\,(\rho-\rho_g)^2 & \textup{for} \;\;\rho<\rho_c
\end{array}
\right.
\end{equation}
and the relation between $S(z;q)$ and $G(0,z;q)$ simplifies to 
 \begin{equation}
S(z;q)\;=\;S_b(q)+\frac{\Delta\rho}{\rho'(0)}\,\frac{\;G(0,z;q)\;}{1+q^2\xi^2}
\label{SDP}
\end{equation}
which, like the result (\ref{m4S}) for the Landau quartic potential, does not display any resonance. These, however, are the only models (together with piece-wise combinations of them) for which resonances are entirely absent.

\subsubsection{Lower Bound for $S(0;q)$}

Using the spectral expansion (\ref{spectral}), we note that 
\begin{equation}
\frac{G(0,0;q)-G(0,0;\sqrt{n^2-1}\,\kappa)}{1-\frac{q^2\xi^2}{n^2-1}}\;=\;\sum_{m=0}^\infty\;\frac{|\psi_m(0)|^2}{(E_m+q^2)(\frac{E_m\xi^2}{n^2-1}+1)}
\end{equation}
where the sum only contains even eigenstates, since all odd eigenfunctions vanish at the origin. Using only the ground-state contribution, we observe that
\begin{equation}
\frac{\;G(0,0;q)-G(0,0;\sqrt{n^2-1}\,\kappa)\;}{1-\frac{q^2\xi^2}{n^2-1}}\;>\;\frac{\rho'(0)^2}{\sigma q^2}
\end{equation}
and substitution into (\ref{SRMS}) leads to
\begin{equation}
\frac{S(0;q)}{S_b(q)} \;>\; 1 \;+\;\frac{\Delta\rho\,\rho'(0)}{\sigma q^2\, S_b(0)}
\label{Stoybound}
\end{equation}
showing that, for all single-resonance models, the approximation (\ref{Sapp}) is, in fact, a lower-bound for $S(0;q)$.

\subsection{Trigonometric Model}

While the previous models contained one resonance, or none at all, we present here a model that contains an infinite number of resonances. These occur at values $q=\sqrt{n^2-1}\,\kappa$ but only for even values of $n$, with the odd values missing, since $\sigma_1=\sigma_3=\sigma_5=\cdots=\infty$.
The potential is given by
\begin{equation}\label{TM}
\Delta\phi(\rho)\;=\;\frac{\kappa^2\,\Delta\rho^2}{2\,\pi^2}\,\sin^2\left(\frac{\pi\,(\rho-\rho_g)}{\Delta\rho}\right)
\end{equation}
for $\rho_g\le\rho\le\rho_l$ (see Fig.\ref{Fig1}). The absence of odd terms in the Taylor's expansion of this potential about $\rho_g$ leads to the absence of the odd resonances. 
Solution of the Euler-Lagrange equation (\ref{EL}) determines that the density profile is given by
\begin{equation}
\rho(z)\;=\;\rho_g+\frac{2\,\Delta\rho}{\pi}\,\arctan\,e^{\kappa z}
\end{equation}
leading to a surface tension
\begin{equation}
\sigma\;=\;\frac{2\,\kappa\,\Delta\rho^2}{\pi^2}
\end{equation}
For this model, the OZ equation (\ref{OZGTG}) becomes
\begin{equation}
\big(-\partial^2_z+q^2+\kappa^2-2\kappa^2\,\text{sech}^2(\kappa z)\big)\,G(z,z';q)\;=\;\delta(z-z')
\end{equation}
and has a similar solution to the Landau quartic potential. Ordering the particles so that $z\ge z'$, we find that
\begin{equation}\label{TMG}
G(z,z';q)\;=\;\frac{\;\,e^{-\kappa_q (z-z')}\,}{2\,\xi^2q^2\,\kappa_q\,}\;\big(\xi\,\kappa_q+\tanh(\kappa z)\big)\big(\xi\,\kappa_q-\tanh(\kappa z')\big)
\end{equation}
which, at the origin, takes the value
\begin{equation}\label{TMG00}
G(0,0;q)\;=\;\frac{\kappa_q}{2\,q^2}
\end{equation}

Setting $z'=0$ and integrating (\ref{TMG}) over $z$ determines the value of the local structure factor at the origin:
\begin{equation}\label{TMS0}
S(0;q)\;=\;\frac{1}{q^2}\left(1+\int_0^\infty\!\!\!\!dt\;\;e^{-\sqrt{1+\xi^2q^2}\;t}\;\tanh(t)\right)
\end{equation}
Although this integral does not have an analytic solution for all $q$, it can be evaluated exactly at the resonances. For example, $S(0,\sqrt{3}\,\kappa)=(\frac{1}{2}+\log 2)\xi^2/3$, and $S(0,\sqrt{8}\,\kappa)=(\frac{1}{3}-\frac{\pi}{16})\xi^2$.

\subsection{Double-Cubic Model}

Our final example of an integrable system corresponds to the potential for the double-cubic model, given by
\begin{equation}\label{DC}
      \Delta\phi(\rho)=
\left\{
\begin{array}{ll}
 \displaystyle\frac{\kappa^2}{2}\,(\rho-\rho_l)^2 \left(1+\frac{4}{3}\,\frac{\rho-\rho_l}{\Delta\rho}\right) & \textup{for} \;\;\rho>\rho_c\\[.5cm]
 \displaystyle\frac{\kappa^2}{2}\,(\rho-\rho_g)^2 \left(1-\frac{4}{3}\,\frac{\rho-\rho_g}{\Delta\rho}\right) & \textup{for} \;\;\rho<\rho_c
\end{array}
\right.
\end{equation}
which satisfies $\phi'(\rho_c)=0$, as required (see Fig.\ref{Fig1}). Unlike the previous models, this contains all resonances.
The density profile follows from solution of the Euler-Lagrange equation (\ref{EL}):
\begin{equation}
      \rho(z)=
\left\{
\begin{array}{ll}
 \rho_l-\displaystyle\frac{3}{4}\,\Delta\rho\;\textup{sech}^2\,\frac{\,\kappa(z-z_0)}{2} & \textup{for} \;\;z>0\\[.5cm]
 \rho_g+\displaystyle\frac{3}{4}\,\Delta\rho\;\textup{sech}^2\,\frac{\,\kappa(z+z_0)}{2} & \textup{for} \;\;z<0
\end{array}
\right.
\end{equation}
with $e^{\kappa z_0}=(\sqrt{3}-1)/(\sqrt{3}+1)$ or, equivalently, $\tanh(-\kappa z_0/2)=1/\sqrt{3}$. The surface tension, given by (\ref{st}), evaluates as:
\begin{equation}
\sigma\;=\;\frac{9-2\sqrt{3}}{30}\;\kappa\,\Delta\rho^2
\end{equation}
which is smaller than the value of $\sigma_1$ determined by (\ref{sigma1}), $\sigma_1=\kappa\,\Delta\rho^2/4$.

The Ornstein-Zernike equation for the correlation function reads
\begin{equation}
\left(-\partial_z^2+\kappa^2+q^2-3\,\kappa^2\textup{sech}^2\frac{|z|-z_0}{2}\right)\;G(z,z';q)\;=\;\delta(z-z')
\end{equation}
which can be solved using the same methods described above. If $z\ge z'\ge 0$, we find that
\begin{equation}
G(z,z';q)\;=\;\alpha(q)\;\psi_-(z;q)\,\psi_+(z';q)\;+\;\beta(q)\;\psi_-(z;q)\,\psi_-(z';q)
\end{equation}
while, if $z\ge 0\ge z'$,
\begin{equation}
G(z,z';q)\;=\;\gamma(q)\;\,\overline{\psi}_+(z';q)\,\psi_-(z;q)
\end{equation}
where
\begin{equation}
      \begin{array}{ccl@{\,}l}
\psi_-(z;q) & = & e^{-\kappa_q z} & \Big(\tau^3+2\,\xi\,\kappa_q\tau^2+(1+\frac{8}{5}\,\xi^2q^2)\tau+\frac{8}{15}\,\xi^3\kappa_q q^2\Big)\\[.35cm]
\psi_+(z;q) & = & e^{\kappa_q z} & \Big(\tau^3-2\,\xi\,\kappa_q\tau^2+(1+\frac{8}{5}\,\xi^2q^2)\tau-\frac{8}{15}\,\xi^3\kappa_q q^2\Big)\\[.35cm]
 \overline{\psi}_+(z;q) & = & e^{\kappa_q z} & \Big(\bar{\tau}^3-2\,\xi\,\kappa_q\bar{\tau}^2+(1+\frac{8}{5}\,\xi^2q^2)\bar{\tau}-\frac{8}{15}\,\xi^3\kappa_q q^2\Big)
      \end{array}
\end{equation}
and
\begin{equation}
\tau=\tanh\frac{\kappa(z\!-\!z_0)}{2},\hspace{1cm}\bar{\tau}=\tanh\frac{\kappa(z\!+\!z_0)}{2},\hspace{1cm}\kappa_q=\sqrt{\kappa^2+q^2}
\end{equation}
The amplitudes of the correlation function are determined by
 \begin{equation}
\begin{array}{ccl}
\alpha(q)&=&\displaystyle\frac{225}{\,\xi^2q^2\kappa_q(5-4\,\xi^2q^2)(3+4\,\xi^2q^2)}\\[.5cm]
\beta(q)&=&\displaystyle\frac{225\,(5+4\,\xi^2q^2)}{4\,\xi^2q^2\kappa_q(5-4\,\xi^2q^2)(3+4\,\xi^2q^2)\epsilon(q)}\\[.5cm]
 \gamma(q)&=&-\displaystyle\frac{225}{8\,\xi\,q^2\epsilon(q)}
\end{array}
\end{equation}
where
 \begin{equation}
 \epsilon(q)=\left(5+4\,\xi^2q^2+\frac{12}{\sqrt{3}}\,\xi\,\kappa_q\right)\left(\frac{10}{\sqrt{3}}+\frac{12}{\sqrt{3}}\,\xi^2q^2+\xi\,\kappa_q(5+4\,\xi^2q^2)\right)
 \end{equation}
is a smooth monotonically increasing function of $q$ that displays no divergences. \\

\noindent
With one particle at the origin, the correlation function on the liquid side ($z\ge0$) reduces to
   \begin{equation}
G(0,z;q)\;=\;\frac{\tau^3+2\,\xi\,\kappa_q\tau^2+(1+\frac{8}{5}\,\xi^2q^2)\tau+\frac{8}{15}\,\xi^3q^2\kappa_q}{4\,\xi\,q^2\Big(\frac{1}{3}+\frac{4\sqrt{3}}{15}\,\xi\,\kappa_q+\frac{4}{15}\,\xi^2q^2\Big)}\;\,e^{-\kappa_q z}
\label{DCG0}
\end{equation}
so that
   \begin{equation}
G(0,0;q)\;=\;\frac{\frac{4\sqrt{3}}{3}+\frac{8\sqrt{3}}{5}\,\xi^2q^2+2\,\xi\,\kappa_q+\frac{8}{5}\,\xi^3q^2\kappa_q}{4\,\xi\,q^2\Big(1+\frac{4\sqrt{3}}{5}\,\xi\kappa_q+\frac{4}{5}\,\xi^2q^2\Big)}
\label{DCG00}
\end{equation}

Integration of $G(0,z;q)$ determines $S(0;q)$ exactly 
\begin{equation}
      S(0;q)=\frac{\;\frac{2}{3}+\frac{4}{15}\,\xi^2q^2+\sqrt{\frac{1+\xi^2q^2}{3}}-\frac{1}{5}\,q^2\xi\,J(q)\;}{q^2\left(\frac{1}{3}+\frac{4}{5}\sqrt{\frac{1+\xi^2q^2}{3}}+\frac{4}{15}\,\xi^2q^2\right)}
\end{equation}
where
\begin{equation}
      J(q)=\int_0^\infty\!\!\!dz\;\;e^{-\sqrt{\kappa^2+q^2}\,z}\,\tanh\left(\frac{\kappa(z\!-\!z_0)}{2}\right)
\end{equation}
is itself a hyper-geometric function. Again, this integral cannot be solved analytically, except at the resonances.

\section{Comparison with Approximations}

These analytical results allow us to test the robustness of the approximations (\ref{Sapp}) and (\ref{Gapp}).
Let us consider the models one by one.\\

{\bf A) Landau Quartic Model}. In this case, as mentioned above, both approximations recover the exact results.\\

{\bf B) Single-Resonance Model $n=2$}. In order to compare the exact result for $G(0,0;q)$, Eq.~(\ref{SRMG00}), with the approximation (\ref{Gapp}), we can rewrite the former as
\begin{equation}\label{SRMcorr}
\frac{G(0,0;q)}{G_b(0;q)}\;=\;\, 1\;+\;\frac{\rho'(0)^2}{\sigma q^2G_b(0;0)}\;\mathcal{C}(q)
\end{equation}
where the correction term $\mathcal{C}(q)$ would be $1$ if the approximation was exact. Instead, we obtain
\begin{equation}
\mathcal{C}(q)\;=\;\frac{2+3\sqrt{2}+(3+\sqrt{2})\sqrt{1+\xi^2 q^2}}{3+3\sqrt{2}+(2+\sqrt{2})\sqrt{1+\xi^2 q^2}}\;\;\approx\;\; 1\;+\;\frac{4\sqrt{2}-5}{14}\,\xi^2q^2+\cdots
\end{equation}
which is a smooth function that ranges between $1$, for $q=0$, and $(4-\sqrt{2})/2\approx 1.293$, for $q\to\infty$. For example, at the resonance, we have $\mathcal{C}(\sqrt{3}\,\kappa)=(8+5\sqrt{2})/(7+5\sqrt{2})\approx 1.071$. Note, however, that the term containing the correction $\mathcal{C}(q)$ in (\ref{SRMcorr}) becomes progressively less important as $q$ increases, indicating that the approximation (\ref{Gapp}) is very robust over the whole range of wave vectors.\\

\begin{figure}[t]
\includegraphics[width=\columnwidth]{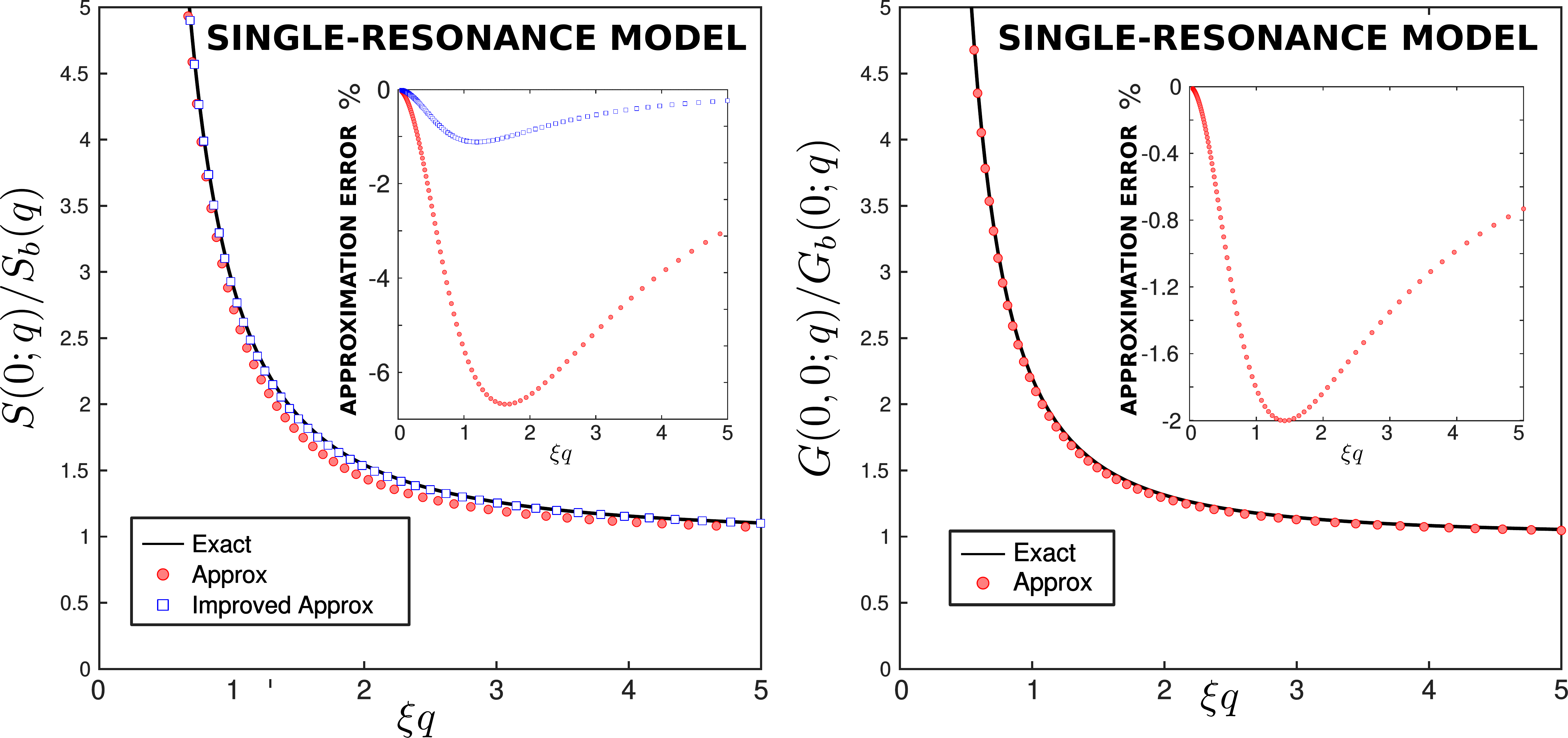}
\caption{\label{Fig2} Comparison of the exact expressions (\ref{SRMS0}) and (\ref{SRMG00}) for $S(0;q)$ and $G(0,0;q)$ with the general approximations (\ref{Sapp}) and (\ref{Gapp}) for the model with a single resonance at $q=\sqrt{3}\,\kappa$. Percentage errors are shown in the insets. For the local structure factor, comparison is also made with the improved approximation (\ref{SappImp}). See text for details.
}
\end{figure}

We now turn our attention to the local structure factor $S(0;q)$. We have already noted that the approximation (\ref{Sapp}) is a lower bound for all single-resonance models. It also contains the correct Goldstone mode divergences as $q\to 0$, and the correct bulk limit as $q\to\infty$, remaining an accurate approximation over the whole range of wave vectors. For example, at the resonance, the inequality (\ref{Stoybound}) implies that $S(0;\sqrt{3}\kappa)= 0.414 \,\xi^2>0.386\,\xi^2$ implying that the lower bound on the RHS is only $7\%$ off the exact result. 
Recall that the approximation (\ref{Sapp}) is obtained by substituting $G(0,0;q)\approx\rho'(0)^2/\sigma q^2 + \mathcal{C}$ into the general result (\ref{Sres}). For the single-resonance model, however, we know that all $\sigma_n$ are infinite except for $\sigma_2=\sigma$. We can, therefore, substitute the improved approximation (\ref{Gapp}) into (\ref{SRMSzq}) to obtain an 
\begin{equation}
S(0;q)\;\approx\; S_b(q)\,\mathcal{A}(q)+\frac{\Delta\rho\,\rho'(0)}{\sigma q^2(1+\xi^2q^2)}\;\mathcal{B}(q)
\label{SappImp}
\end{equation}
where 
\begin{equation}
\mathcal{A}(q)\;=\;1+\frac{3}{\sqrt{2}(2+\sqrt{1+q^2\xi^2})\sqrt{1+q^2\xi^2}}
\end{equation}
and
\begin{equation}
\mathcal{B}(q)\;=\;\frac{1+\frac{1}{3}q^2\xi^2+\frac{1}{12}q^4\xi^4}{(1+\frac{1}{6}q^2\xi^2\sqrt{1+q^2\xi^2})\sqrt{1+q^2\xi^2}}
\end{equation}
Fig.~\ref{Fig2} shows the comparison between the exact results and the approximations discussed here.\\

{\bf C) Trigonometric Model}.  For this model, as with the Landau quartic potential, the approximation (\ref{Gapp}) for $G(0,0;q)$ is exact, and recovers the analytical result (\ref{TMG00}) identically. In fact, within this model, there is a remarkably simple result for the correlation function even away from the interface. Setting $z=z'$ in (\ref{TMG}), we find that 
\begin{equation}
G(z,z;q)\;=\; \frac{1}{2\,\kappa_q} \left(1+\frac{\text{sech}^2(\kappa\,z)}{q^2\,\xi^2}\right)
\end{equation}
which can be rewritten
\begin{equation}
G(z,z;q)\;=\;\frac{1}{2\kappa_q}+\frac{\rho'(z)^2}{\sigma q^2\sqrt{1+q^2\xi^2}}
\end{equation}
or, equivalently
\begin{equation}\label{TRGzz}
\frac{G(z,z;q)}{G_b(0;q)}\;=\;1+\frac{\rho'(z)^2}{\sigma q^2\,G_b(0;0)}
\end{equation}

The correlation function $G(z,z;q)$ takes its maximum value at the origin (where $\rho'(z)$ is maximum), and expression (\ref{TRGzz}) is therefore equivalent to (\ref{Gapp}). As $|z|$ increases, the Goldstone mode contribution decreases, and $G(z,z;g)$ approaches its bulk limit $G_b(0;q)$.

\begin{figure}[t]
\includegraphics[width=0.5\columnwidth]{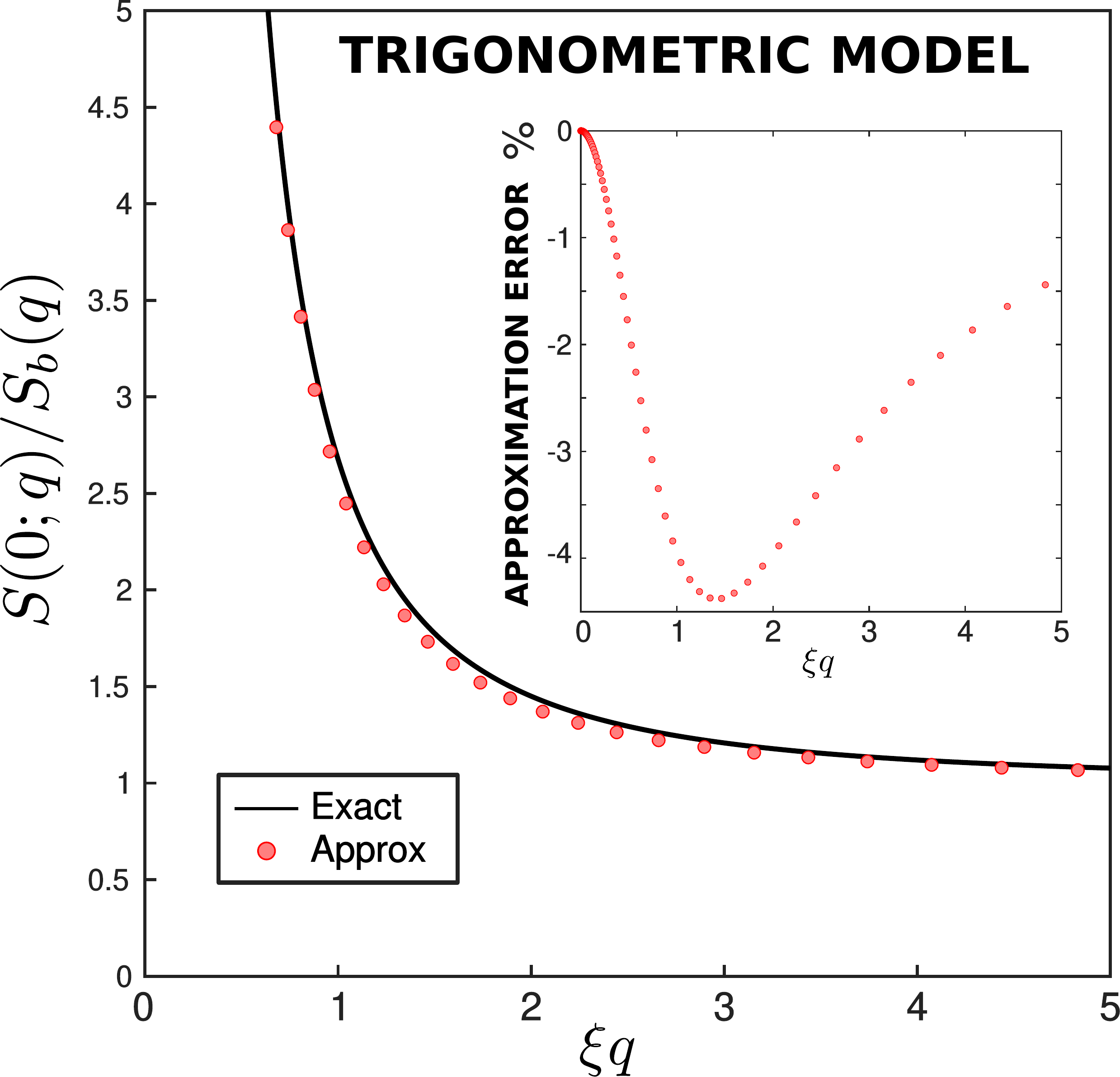}
\caption{\label{Fig3} Comparison of the exact expression (\ref{TMS0}) for $S(0;q)$ with the general approximation (\ref{Sapp}) for the trigonometric model.  Percentage errors are shown in the inset. For this model, the approximation (\ref{Gapp}) for $G(0,0;q)$ recovers the exact analytical result.
}
\end{figure}

The approximation for $S(0;q)$ is not exact, but again remains accurate over the whole range of vectors. For example, at the first resonance, $S(0;\sqrt{3}\,\kappa)=0.3906\,\xi^2$, while the approximation (\ref{Sapp}) yields $S(0;\sqrt{3}\,\kappa)\approx 0.3803\,\xi^2$. See Fig.\ref{Fig3}.\\

{\bf D) Double-Cubic Model}. Following our analysis of the single-resonance model, we rewrite the result in the form (\ref{SRMcorr}), allowing for a correction term $\mathcal{C}(q)$, which evaluates as
\begin{equation}
\mathcal{C}(q)\;=\;(9-2\sqrt{3})\;\frac{1+\frac{2\sqrt{3}}{3}\sqrt{1+\xi^2 q^2}+\frac{4}{5}\,\xi^2 q^2}{5+4\sqrt{3}\sqrt{1+\xi^2 q^2}+4\,\xi^2 q^2}\;\;\approx\;\; 1\;+\;\frac{39\sqrt{2}-66}{115}\,\xi^2q^2+\cdots
\end{equation}

\begin{figure}[t]
\includegraphics[width=\columnwidth]{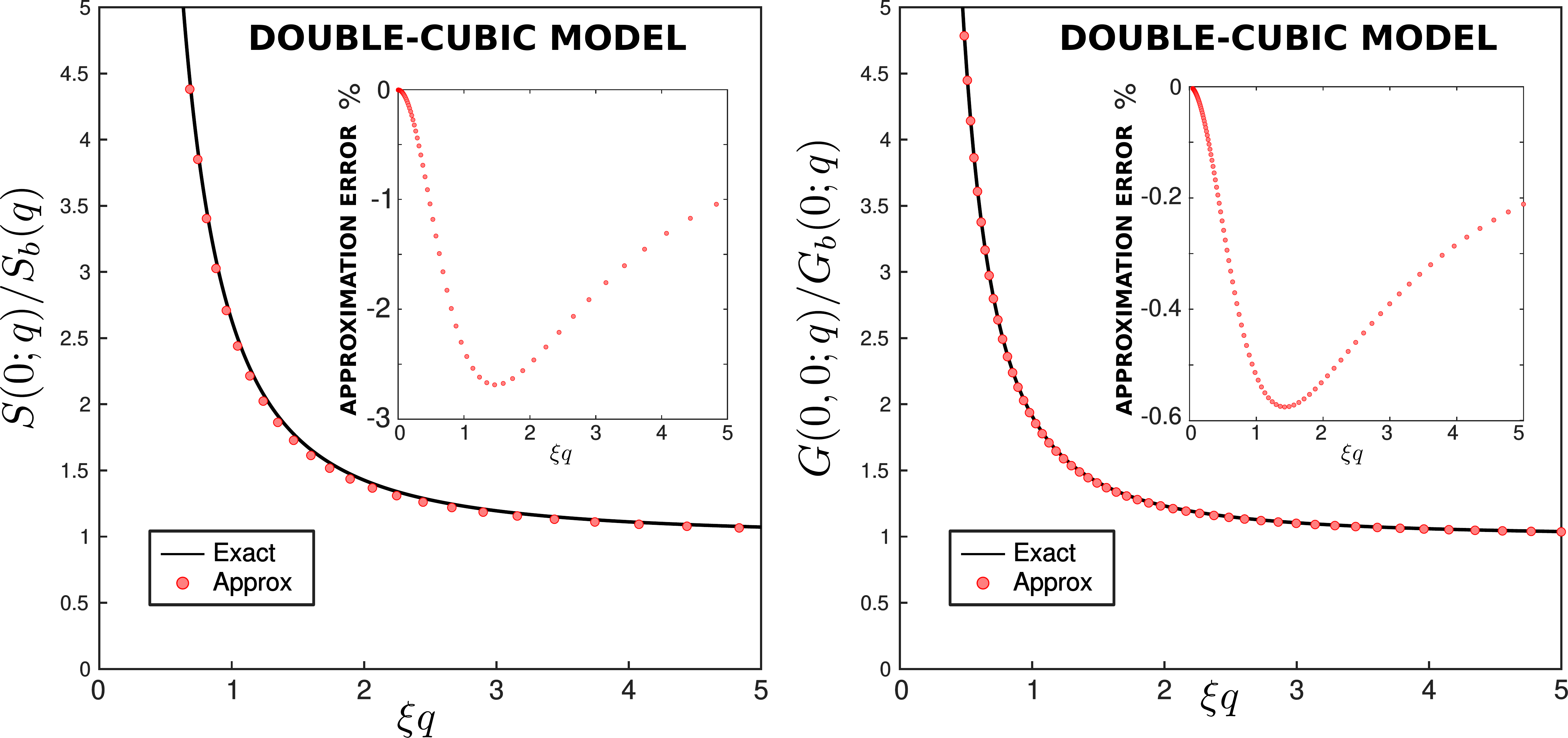}
\caption{\label{Fig4} Comparison of the exact expressions (\ref{SRMS0}) for $S(0;q)$ and (\ref{SRMG00}) for $G(0,0;q)$ with the general approximations (\ref{Sapp}) and (\ref{Gapp}) for the double-cubic model. Percentage errors are shown in the insets.
}
\end{figure}

This correction term is even smaller than for the single-resonance model. It ranges between $1$, for $q=0$, and $(9-2\sqrt{3})/5\approx 1.107$, for $q\to\infty$, and, at the first resonance, $\mathcal{C}(\sqrt{3}\,\kappa)\approx 1.024$. This reflects the fact that the coefficient of $q^2$ in the small $q$ expansion is near negligible, taking the value $0.01348\,\xi^2$. As seen in Fig.~\ref{Fig4}, the maximum error is only about $0.6\%$ for $G(0,0;q)$.\\

The approximation for the local structure factor at the origin (\ref{Sapp}) is also accurate: It is exact in the limits $q\to 0$ and $q\to\infty$, with a maximum error of about $2.5\%$. See Fig.\ref{Fig4}. \\

\begin{figure}[t]
\includegraphics[width=\columnwidth]{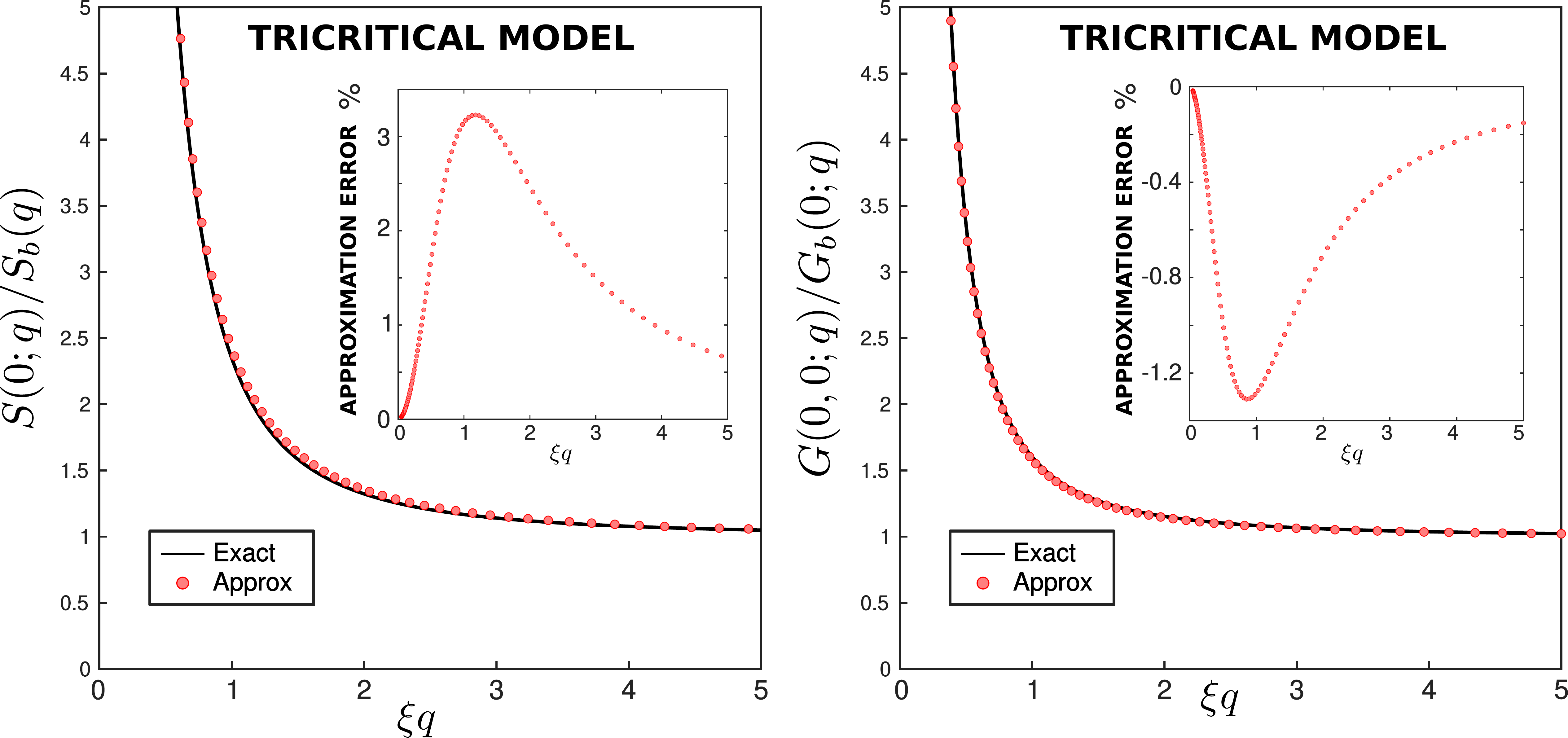}
\caption{\label{Fig5} Comparison of the general approximations (\ref{Sapp}) and (\ref{Gapp}) for $S(0;q)$ and $G(0,0;q)$ with the results obtained from solving numerically the OZ equations (\ref{OZGTS}) and  (\ref{OZGTG}) for the tricritical model. Percentage errors are shown in the insets.
}
\end{figure}

{\bf E) Tricritical Model}. As a final test of (\ref{Sapp}) and (\ref{Gapp}), we turn attention to the model potential
\begin{equation}
\phi(\rho)=-\frac{t}{2} (\rho-\rho_c)^2+\frac{u}{6}(\rho-\rho_c)^6
\label{FW}
\end{equation}
or, equivalently,
\begin{equation}
\Delta\phi(\rho)\;=\;\frac{\;\kappa^2}{2}(\rho-\rho_l)^2\;\left(1+\frac{10}{3}\;\frac{\rho-\rho_l}{\Delta\rho}+5\,\left(\frac{\rho-\rho_l}{\Delta\rho}\right)^2+4\,\left(\frac{\rho-\rho_l}{\Delta\rho}\right)^3+\frac{4}{3}\,\left(\frac{\rho-\rho_l}{\Delta\rho}\right)^4\right)
\end{equation}
describing the approach to a bulk tricritical point, occurring at $t=0$ (see Fig.\ref{Fig1}) \cite{Rowlinson1982}. The upper critical dimension for bulk tricriticality is $d^*=3$, and the mean-field predictions $\Delta\rho\propto t^\frac{1}{4}$ and $\kappa\propto t^\frac{1}{2}$ remain valid in three dimensions up to minor corrections. From (\ref{sigma1}), it follows that
\begin{equation}
\sigma_1=\frac{\kappa\Delta\rho^2}{10}
\end{equation}
which is smaller than the surface tension $\sigma\approx \kappa\Delta\rho^2/7$, obtained from numerical integration of (\ref{st}). Clearly, both $\sigma_1$ and $\sigma$ vanish $\propto t$ approaching the tricritical point but with different amplitudes. This means that the remaining weights $\sigma_n/\sigma$ do not vanish as $t\to 0$, implying that the resonances are fully present (in contrast to the standard quartic potential describing the approach to bulk criticality (\ref{m4})). Fig.~\ref{Fig5} compares the approximations (\ref{Sapp}) and (\ref{Gapp}) for $S(0;q)$ and $G(0,0;q)$ with those obtained from  numerical solution of the Ornstein-Zernike equations (\ref{OZGTS}) and (\ref{OZGTG}) and again demonstrate their extraordinary accuracy and utility over the whole range of wave-vectors. For example, the approximate expression for $G(0,0;q)/G_b(q)$, which recall is exact at low and high $q$, is only $1\%$ inaccurate, at worst, and is barely indistinguishable from the exact numerical result. Finally, we note that, on simply replacing $t$ with $t^\frac{4}{3}$, the same potential (\ref{FW}) can be viewed as a phenomenological Fisk-Widom model describing the approach to a bulk critical point with the rational approximations for the three dimensional critical singularities $\Delta\rho\propto t^\frac{1}{3}$,  $\kappa\propto t^\frac{2}{3}$ and $\sigma\propto t^\frac{4}{3}$ \cite{Fisk1969}.

\section{Conclusions}

\begin{table}
\begin{tabular}{|c|c|c|}
\multicolumn{1}{c}{}  & \multicolumn{1}{c}{\large$S(0;q)$} & \multicolumn{1}{c}{\large$G(0,0;q)$} \\[.1cm]
\hline 
& & \\[-.25cm]
Landau Quartic & \hspace{.5cm}EXACT\hspace{.5cm} & \hspace{.5cm}EXACT\hspace{.5cm} \\ 
& & \\[-.25cm]
\hline 
& & \\[-.25cm]
\hspace{.15cm}Single-Resonance Model\hspace{.15cm} & $-6.7\%$ & $-2.0\%$ \\ 
& & \\[-.25cm]
\hline 
& & \\[-.25cm]
Trigonometric & $-4.4\%$ & EXACT \\ 
& & \\[-.25cm]
\hline 
& & \\[-.25cm]
Double-Cubic & $-2.7\%$ & $-0.58\%$ \\ 
& & \\[-.25cm]
\hline 
& & \\[-.25cm]
Tricritical & $3.2\%$ & $-1.3\%$ \\[.1cm]
\hline 
\end{tabular}
\caption{Summary of the maximum relative errors of the approximations (\ref{Sapp}) for $S(0;q)$ and (\ref{Gapp}) for $G(0,0;q)$, when compared with the exact results for the five models shown in Figs.~\ref{Fig2} to \ref{Fig5}.}\label{Table1}
\end{table}

In this paper, we have used square-gradient theory to test the robustness of approximations for the microscopic structure factor $S(0;q)$  and correlation function $G(0,0;q)$ at the liquid-gas interface which emerge from an analysis of the resonances occurring in the tails of $S(z;q)$. Comparison with new analytical and numerical results obtained from solution of the Ornstein-Zernike equation shows the remarkable accuracy of these approximations, which are indeed exact for several model potentials $\phi(\rho)$. A summary is given in Table \ref{Table1}. While the results presented here are specific to simple square-gradient theory, almost identical results apply to the more microscopic Sullivan model of the interfacial region \cite{Parry2019}. These results demonstrate further that for systems with short-ranged forces one can essentially determine analytically the microscopic correlation function structure in the interfacial region, in both density-functional theory and simulation studies, without resorting to mesoscopic concepts such as a wave-vector dependent surface tension.\\

To finish our paper, we make a number of remarks. Firstly, we have tested if it is possible to improve upon the approximation (\ref{Sapp}) for $S(0;q)$, which, recall, arises when we approximate $G(0,0;q)\approx\rho'(0)^2/\sigma q^2 + \mathcal{C}$ in the resonant expansion (\ref{Sres}). For example, one may attempt to do this by using the better approximation (\ref{Gapp}) for the correlation function, and substitute it into the general result (\ref{Sres}). Indeed, we showed that this is possible for the single-resonance model where we know a priori that $\sigma_2=\sigma$. However, in general, the values of the coefficients $\sigma_n$ for $n\ge 2$ are not known and its is necessary to make further approximations. We may, for instance, include only the first resonance at $q=\sqrt{3}\kappa$ and set $1/\sigma_2=1/\sigma- 1/\sigma_1$ to ensure that the correct Goldstone mode divergence is recovered identically in the limit $q\to 0$. We have investigated this and shown that this only improves the accuracy for some model potentials. For the double-cubic model, it reduces the maximum error in $S(0;q)$ from $-6.7\%$ to about $0.088\%$. However, the same approximation does not significantly improve the accuracy of $S(0;q)$ for the trigonometric model and worsens it for the tricritical potential. We must conclude, therefore, that the simple approximation (\ref{Sapp}) is the most robust across all model potentials.

Secondly, we note that there are further examples of fully integrable square-gradient theories for which the correlation function $G(0,0;q)$ can be determined analytically. Indeed, there is an infinite class of these, corresponding to models for which the density profile satisfies $\rho'(z)\propto \text{sech}^N (\kappa z/N)$. The values $N=1$ and $N=2$ correspond respectively to the trigonometric and Landau quartic potentials where, recall, the result (\ref{Gapp}) for $G(0,0;q)$ is exact. The correlation function $G(z,z';q)$ for all these models can be determined exactly and is of the same form (\ref{m4Gzz}). Here, the function $\psi(z;q)=e^{-\kappa_q z}P(y)$, where $P(y)$ is a Jacobi polynomial of degree $N$ in $y=\tanh ({\kappa z/N})$. For $N>2$, the result (\ref{Gapp}) is no longer exact but can be recast in the form (\ref{SRMcorr}), including a small correction term $C(q)$. For example, for $N=3$, we find $C(q)=(1+\frac{6}{5}q^2\xi^2)(1+\frac{9}{8}q^2\xi^2)$, which makes a near negligible correction to the very accurate approximation (\ref{Gapp}). These generalised integrable models correspond to potentials $\phi(\rho)$ which, except for $N=1,2$, do not have an analytic expansion about the bulk densities, and have density profiles whose expansions are not analytic in $X=e^{-\kappa z}$. Thus, although these models lie outside the classification of potentials for which the resonant expansion (\ref{Sres}) applies, the unexpected agreement further testifies to the robustness of the approximation (\ref{Gapp}).

Finally, we mention that here we have only considered model systems that display an Ising symmetry. Of course, more generally, a certain degree of asymmetry is to be expected, and the liquid and gas phases are characterised by distinct bulk correlation lengths. For the local structure factor, this issue is not difficult to address and has been discussed in \cite{Parry2019}, where it was shown that the approximation (\ref{Sapp}) generalises in a straightforward manner. Indeed, this  was used to capture the local structure factor for the Sullivan model near perfectly using an accurate Carnahan-Starling equation of state. The integrable models described here provide a means of studying how the result (\ref{Gapp}) for the correlation function generalises when liquid-gas asymmetry is allowed for, and will be discussed in a following paper.

\acknowledgments

AOP acknowledges the EPSRC, UK for grant EP/L020564/1 (Multiscale analysis of complex interfacial phenomena). CR acknowledges the support of the grant PGC2018-096606-B-I00 (MCIU/AEI/FEDER, UE).

\bibliography{wetting}

\pagebreak

\end{document}